\documentstyle[aps,12pt]{revtex}
\begin{document}
\title{Transitional Lu and Spherical Ta Ground-State Proton 
Emitters in the Relativistic Hartree-Bogoliubov model}
\author{G.A. Lalazissis$^{1}$, D. Vretenar$^{1,2}$, and P. Ring$^{1}$}
\address{
$^{1}$ Physik-Department der Technischen Universit\"at M\"unchen,
D-85748 Garching, Germany\\
$^{2}$ Physics Department, Faculty of Science, University of
Zagreb, Croatia\\}
\maketitle
\bigskip
\bigskip
\begin{abstract}
Properties of transitional Lu and spherical Ta ground-state
proton emitters are calculated with the Relativistic Hartree Bogoliubov
(RHB) model. The NL3 effective interaction is used in the mean-field 
Lagrangian, and pairing correlations are described by the 
pairing part of the finite range Gogny interaction D1S.
Proton separation energies, ground-state quadrupole
deformations, single-particle orbitals occupied by the odd
valence proton, and the corresponding spectroscopic factors
are compared with recent experimental data, and with results 
of the macroscopic-microscopic mass model.
\end{abstract}
\vspace{1 cm}
\vspace{1 cm}\\
The structure and decays modes of nuclei beyond the proton drip-line
represent one of the most active areas of experimental and theoretical 
studies of exotic nuclei with extreme isospin values. In the last
few years many new data on ground-state and isomeric proton 
radioactivity have been reported. In particular, detailed studies
of odd-Z ground-state proton emitters in the regions 
51$\leq $Z$\leq $55 and 69$\leq $Z$\leq $ 83, have shown that the 
systematics of spectroscopic factors is consistent with
half-lives calculated in the spherical WKB or distorted-wave Born 
(DWBA) approximations
\cite{WD.97,ASN.97}. More recent data~\cite{Dav.98,Ryk.99} 
indicate that the missing region of light rare-earth nuclei contains
strongly deformed systems at the drip-lines.

In the theoretical description of ground-state and isomeric proton
radioactivity, two essentially complementary approaches have been 
reported. One possibility is to start from a spherical or deformed
phenomenological single-particle potential, a Woods-Saxon potential
for instance, and to adjust the parameters of the potential well
in order to reproduce the experimental one-proton separation energy.
The width of the single-particle resonance is then determined by
the probability of tunneling through the Coulomb and centrifugal 
barriers. Since the probability strongly depends on the valence
proton energy and on its angular momentum, the calculated half-lives
provide direct information about the spherical or deformed orbital
occupied by the odd proton. For a spherical proton emitter it is
relatively simple to calculate half-lives in the WKB
or DWBA approximations~\cite{ASN.97}. On the other hand, it is 
much more difficult to quantitatively describe the process of 
three-dimensional quantum mechanical tunneling for deformed
proton emitters. Modern reliable models for calculating 
proton emission rates from deformed nuclei have been developed
only recently~\cite{Ryk.99,MFL.98}. A shortcoming of this approach
is that it does not predict proton separation energies, i.e.
the models do not predict which nuclei are likely to be 
proton emitters. In fact, if they are used to calculate 
decay rates for proton emission from excited states, the 
depth of the central potential has to to be adjusted for 
each proton orbital separately. In addition, the models
of Refs.~\cite{ASN.97,Ryk.99,MFL.98} do not provide any 
information about the spectroscopic factors of the proton 
orbitals. Instead, experimental spectroscopic factors are
defined as ratios of calculated and measured half-lives, 
and the deviation from unity is attributed to nuclear 
structure effects. 

In Refs.~\cite {VLR.98,LVR.99,VLR.99} we have used the 
relativistic Hartree Bogoliubov (RHB) theory to calculate 
properties of proton-rich spherical even-even nuclei with 
14$\leq $Z$\leq $28, and to describe odd-Z deformed ground-state
proton emitters in the region $53 \leq Z \leq 69$. RHB presents
a relativistic extension of the Hartree-Fock-Bogoliubov theory,
and it provides a unified framework for the description of
relativistic mean-field and pairing correlations.Such a unified and
self-consistent formulation is especially important in
applications to drip-line nuclei. The RHB framework 
has been used to study the location of the proton drip-line,
the ground-state quadrupole deformations and one-proton separation
energies at and beyond the drip line, the deformed single particle
orbitals occupied by the odd valence proton, and the corresponding
spectroscopic factors. The results of fully self-consistent calculations
have been compared with experimental data on ground-state proton
emitters. However, since it is very difficult to use the 
self-consistent ground-state wave functions in the calculation
of proton emission rates, one could say that the RHB model provides 
informations which are complementary to those obtained with the models 
of Refs.~\cite{ASN.97,Ryk.99,MFL.98}. 
It should be noted that in the relativistic framework the strength and the
shape of the spin-orbit term are determined selfconsistently. This is
essential for a correct description of spin-orbit splittings in regions 
of nuclei far from stability, where the extrapolation of effective strength
parameters becomes questionable. 
The motivation for the present
work are the very recent data on proton emission from the closed 
neutron shell nucleus $^{155}$Ta~\cite{Uus.99}, and the proposed
experiment to search for direct proton emission
from $^{149}$Lu~\cite{Bat.99}. The analysis of ground-state
proton radioactivity in the Lu and Ta isotopes completes 
our study of deformed and transitional proton emitters in the 
region $53 \leq Z \leq 73$.

A very detailed description of the relativistic Hartree-Bogoliubov 
theory can be found, for instance, in Ref. \cite{LVR.99}. In the following 
we only outline the essential features of the model that will be used
to describe nuclei at the proton drip-line. The ground state of a nucleus
is represented by the Slater determinant of independent 
single-quasiparticle states, which are obtained as solutions of the
relativistic Hartree-Bogoliubov equations 
\begin{eqnarray}
\label{equ.2.2}
\left( \matrix{ \hat h_D -m- \lambda & \hat\Delta \cr
                -\hat\Delta^* & -\hat h_D + m +\lambda} \right)
                \left( \matrix{ U_k({\bf r}) \cr V_k({\bf r}) } \right) =
                E_k\left( \matrix{ U_k({\bf r}) \cr V_k({\bf r}) } \right).
\end{eqnarray}
The column vectors denote the quasi-particle spinors and $E_k$
are the quasi-particle energies. In the Hartree approximation for
the self-consistent mean field, the single-nucleon Dirac
Hamiltonian reads
\begin{equation}
\hat{h}_{D}=-i{\mathbf{\alpha \cdot \nabla }}+\beta (m+g_{\sigma }\sigma
({\mathbf r}))+g_{\omega }\tau _{3}\omega ^{0}({\mathbf r})+
g_{\rho }\rho ^{0}({\mathbf r})+
e{\frac{{(1-\tau _{3})}}{2}}A^{0}({\mathbf r}).  \label{dirh}
\end{equation}
It describes the motion of independent Dirac nucleons in the 
mean-field potentials: the isoscalar scalar $\sigma$-meson potential, 
the isoscalar vector $\omega$-meson, and the isovector vector $\rho$-meson
potential. The photon field $A$ accounts for the electromagnetic
interaction. The meson potentials are determined self-consistently
by the solutions of the corresponding Klein-Gordon equations. 
The source terms for these equations are calculated 
in the {\it no-sea} approximation. Because of charge conservation only the
third component of the isovector $\rho$-meson contributes.
For an even-even system, due to time reversal invariance
the spatial vector components \mbox{\boldmath $\omega,~\rho_3$} and
${\bf  A}$ of the vector meson fields vanish. 
In nuclei with odd numbers of protons or neutrons time
reversal symmetry is broken. The odd particle induces
polarization currents and the time-odd components in the
meson fields. These components play an essential role in the
description of magnetic moments and of moments
of inertia in rotating nuclei. However,
their effect on deformations and binding energies is
very small and can be neglected to a good approximation.
As in our previous studies of nuclei at the proton drip-lines, 
we choose the NL3 set of meson masses and meson-nucleon coupling 
constants~\cite{LKR.97} for the effective interaction
in the particle-hole channel: $m=939$ MeV, $m_{\sigma}=508.194$ MeV,
$m_{\omega}=782.501$ MeV, $m_{\rho}=763.0$ MeV,
$g_{\sigma}=10.217$, $g_2=-10.431$ fm$^{-1}$, $g_3=-28.885$,
$g_{\omega}=12.868$ and  $g_{\rho}=4.474$.

The pairing field $\hat\Delta $ is defined  
\begin{equation}
\label{equ.2.5}
\Delta_{ab} ({\bf r}, {\bf r}') = {1\over 2}\sum\limits_{c,d}
V_{abcd}({\bf r},{\bf r}') {\bf\kappa}_{cd}({\bf r},{\bf r}'),
\end{equation}
where $V_{abcd}({\bf r},{\bf r}')$ are matrix elements of a
two-body pairing interaction, and ${\bf\kappa}_{cd}({\bf r},{\bf r}')$
is the pairing tensor. The pairing part of the phenomenological
Gogny force
\begin{equation}
V^{pp}(1,2)~=~\sum_{i=1,2}
e^{-(( {\bf r}_1- {\bf r}_2)
/ {\mu_i} )^2}\,
(W_i~+~B_i P^\sigma
-H_i P^\tau -
M_i P^\sigma P^\tau),
\end{equation}
with the set D1S \cite{BGG.84} for the parameters
$\mu_i$, $W_i$, $B_i$, $H_i$ and $M_i$ $(i=1,2)$, 
is used to describe pairing correlations. 

The RHB single-quasiparticle equations (\ref{equ.2.2})
are solved self-consistently. The iteration procedure is performed
in the quasi-particle basis. 
The chemical potential $\lambda$  has to be determined by
the particle number subsidiary condition in order that the
expectation value of the particle number operator
in the ground state equals the number of nucleons.
A simple blocking prescription is used in the calculation 
of odd-proton and/or odd-neutron systems.
The blocking calculations are performed without breaking
the time-reversal symmetry. 
The resulting eigenspectrum is
transformed into the canonical basis of single-particle
states, in which the RHB ground-state takes the
BCS form. The transformation determines the energies
and occupation probabilities of the canonical states.

The one-proton separation energies 
\begin{equation}
S_{p}(Z,N) = B(Z,N) - B(Z-1,N)
\label{sep}
\end{equation}
for the Lu and Ta isotopes are displayed in Fig. \ref{figA},
as function of the number of neutrons. The predicted drip-line 
nuclei are $^{154}$Lu and $^{159}$Ta. In the process of proton 
emission the valence particle tunnels through the Coulomb 
and centrifugal barriers, and the decay probability depends
strongly on the energy of the proton and on its angular 
momentum. In rare-earth nuclei the decay of the ground-state by 
direct proton emission competes with $\beta^+$ decay; for heavy nuclei
also fission or $\alpha$ decay can be favored. In general, ground-state
proton emission is not observed immediately after the drip-line.
For small values of the proton separation energies, the width 
is dominated by the $\beta^+$ decay. On the other hand, large 
separation energies result in extremely short 
proton-emission half-lives, which are difficult to observe 
experimentally. For a typical rare-earth nucleus the
separation energy window in which ground-state proton decay can be
directly observed is about 0.8 -- 1.7 MeV \cite{ASN.97}.
In Fig. \ref{figA} we have compared the calculated separation energies 
with experimental transition energies for ground-state
proton emission in $^{150}$Lu, $^{151}$Lu \cite{Sel.93}, $^{155}$Ta
\cite{Uus.99}, $^{156}$Ta \cite{Page.92}, and  $^{157}$Ta \cite{Irv.97}. 
In all five cases an excellent agreement is observed between model 
predictions and experimental data. In addition to $^{151}$Lu, which 
was the first ground-state proton emitter to be discovered~\cite{Hof.82}, 
and $^{150}$Lu, the self-consistent RHB calculation predicts 
ground-state proton decay in $^{149}$Lu. The calculated one-proton
separation energy $-1.77$ MeV corresponds to a half-life of 
a few $\mu s$, if one assumes that the nucleus is spherical. 
Direct proton emission with a half-life of the order of few $\mu s$
is just above the lower limit of observation of current experimental 
facilities. An experiment to search for direct proton emission
from $^{149}$Lu has been proposed recently~\cite{Bat.99}.
For the Lu ground-state proton emitters, in Table \ref{TabA}
the results of the RHB model calculation are compared with the
predictions of the finite-range droplet (FRDM) mass model:
the projection of the odd-proton angular
momentum on the symmetry axis and the parity of the odd-proton state
$\Omega_{p}^{\pi }$ \cite{MNK.97}, the one-proton separation
energy \cite{MNK.97}, and the ground-state quadrupole 
deformation \cite{MN.95}. We have also included the RHB spectroscopic
factors, and compared the separation energies with the experimental
transition energies in $^{150}$Lu and $^{151}$Lu. Both theoretical
models predict oblate shapes for the Lu proton emitters, and similar
values for the ground-state quadrupole deformations.
On the other hand, while the FRDM assigns spin and 
parity $5/2^-$ to the deformed single-particle orbitals occupied by the
odd valence proton in all three proton emitters, the RHB model 
predicts the $7/2^-[523]$ Nilsson orbital to be occupied by the 
odd proton. We also notice that the
RHB separation energies are much closer to the experimental values.
The spectroscopic factors of the $7/2^-[523]$ orbital are displayed 
in the sixth column of Table \ref{TabA}. The spectroscopic factor
of the deformed odd-proton orbital $k$ is defined as the probability
that this state is found empty in the daughter nucleus with even number 
of protons. 

In the detailed analysis of odd-Z proton emitters 
$(53\leq Z\leq 69)$~\cite{LVR.99} it has been shown that,
while the proton-rich isotopes of La, Pr, Pm, Eu and
Tb are all strongly prolate deformed ($\beta _{2}\approx 0.30-0.35$), 
Ho and Tm isotopes at the proton drip-line 
display a transition from prolate to oblate shapes.
Spherical shapes are expected as the nuclei with unbound protons
approach the $N=82$ neutron shell.
The Lu proton emitters are found in the transitional region
between oblate and spherical shapes. This is illustrated in 
Fig.~\ref{figB}, where we plot the binding energy curve for
$^{151}$Lu as function of the quadrupole deformation parameter.
The binding energies result from self-consistent RHB/NL3 
calculations performed by imposing a quadratic constraint
on the quadrupole moment. A very shallow minimum is found
at $\beta \approx -0.15$, but otherwise the potential is 
rather flat with a shoulder at $\beta=0$. In Fig.~\ref{figC}
we compare the ground-state quadrupole deformations of the 
proton-rich Lu isotopes with those of Ho and Tm~\cite{VLR.99}.
For $N \leq 80$ all three chains of isotopes display oblate 
deformations; starting with $N=81$ a sharp transition to the 
spherical shape is observed.

The proton-rich Ta isotopes are spherical. In Fig.~\ref{figA}
we compare the calculated one-proton separation energies with 
experimental transition energies for $^{155}$Ta~\cite{Uus.99},
$^{156}$Ta~\cite{Page.92}, and  $^{157}$Ta~\cite{Irv.97}. 
The predictions for the spherical orbitals occupied by the odd
proton and the corresponding spectroscopic factors are 
displayed in Table \ref{TabB}. Results of the FRDM 
calculation~\cite{MNK.97} have also been included in the comparison. 
As in the case of the Lu ground-state proton emitters,
an excellent agreement between RHB separation energies 
and experimental data on transition energies for proton 
emission is observed. In particular, our calculation reproduces
the very recent data on proton emission from the closed 
neutron shell nucleus  $^{155}$Ta~\cite{Uus.99}. The 
significant decrease in proton binding for $^{155}$Ta, 
as compared to $^{157,156}$Ta, has been associated with the 
$N=82$ closure. In comparison, the FRDM results are found to be in 
rather poor agreement with experimental data.
Except for $^{157}$Ta, the spherical orbitals
predicted to be occupied by the odd proton agree with the 
experimental assignments, and the theoretical spectroscopic 
factor of the $h_{11/2}$ orbital in $^{155}$Ta is very close to 
the experimental value. For $^{157}$Ta the RHB model predicts 
ground-state proton emission from the $h_{11/2}$ orbital. 
The experimental assignment for the ground-state configuration
is $s_{1/2}$, but an alpha decaying state is identified in 
$^{157}$Ta at an excitation energy of only 22(5) keV and 
assigned to an $h_{11/2}$ isomer~\cite{Irv.97}. We have also 
calculated the one-proton separation energy for $^{156}$Ta$^m$:
$S_p = - 1.250$ MeV, the orbital is $h_{11/2}$ and the spectroscopic
factor is 0.79. This is to be compared with the experimental 
transition energy $E_p = 1.103(12)$ MeV~\cite{Liv.93}, assigned to the 
$h_{11/2}$ orbital with the experimental spectroscopic 
factor 0.92(4)~\cite{WD.97}.

In conclusion, the relativistic Hartree-Bogoliubov model has been 
applied in the description of ground-state properties of 
transitional Lu and spherical Ta proton emitters. 
The NL3 effective interaction has been used for the
mean-field Lagrangian, and pairing correlations have been described
by the pairing part of the finite range Gogny interaction D1S.
We would like to emphasize that this particular combination of
effective forces has been used in most of our recent applications
of the RHB model, not only for spherical and deformed
$\beta$-stable nuclei, but also for nuclear systems with
large isospin values on both sides of the valley of $\beta$-stability.
The model parameters therefore have not been adjusted to the specific
properties of nuclei studied in this work, or to the properties of
deformed proton emitters discussed in Refs.~\cite{LVR.99,VLR.99}.
The self-consistent calculation reproduces in detail the
observed transition energies for ground-state
proton emission in $^{150}$Lu, $^{151}$Lu \cite{Sel.93}, $^{155}$Ta
\cite{Uus.99}, $^{156}$Ta \cite{Page.92}, and  $^{157}$Ta \cite{Irv.97},
as well as the assignments for the orbitals occupied by the valence
odd proton and the corresponding spectroscopic factors. 
The model also predicts the one-proton separation energy of $-1.77$ MeV for 
the possible proton emitter $^{149}$Lu. Oblate ground-state deformations
are predicted for all Lu proton emitters, while spherical shapes
are calculated for the Ta isotopes at and beyond the proton drip-line.
With the excellent agreement observed between the RHB results and
the very recent data on proton emission from the closed 
neutron shell nucleus $^{155}$Ta~\cite{Uus.99}, we complete our
study of ground-state properties of deformed and transitional
proton emitters~\cite{LVR.99,VLR.99}.

\newpage
{\bf Figure Captions}
\bigskip

\begin{figure}
\caption{Proton separation energies for Lu and Ta isotopes at and beyond the
drip-line. Results of self-consistent RHB calculations are compared with 
experimental transition energies for ground-state proton emission in 
$^{150}$Lu, $^{151}$Lu \protect\cite{Sel.93}, $^{155}$Ta 
\protect\cite{Uus.99}, 
$^{156}$Ta \protect\cite{Page.92}, and $^{157}$Ta \protect\cite{Irv.97}. 
Filled diamonds 
denote the negative values of the transition energies $E_{p}$.}
\label{figA}
\end{figure}

\begin{figure}
\caption{Potential energy curve for $^{151}$Lu. The 
self-consistent RHB calculations are performed with 
constrained quadrupole deformation.}
\label{figB}
\end{figure}

\begin{figure}
\caption{Self-consistent ground-state quadrupole 
deformations for Ho, Tm and Lu isotopes, at and 
beyond the proton drip-line.}
\label{figC}
\end{figure}

\newpage
\begin{table}
\caption{ Lu ground-state proton emitters. 
Results of the RHB calculation for the one-proton
separation energies $S_p$, quadrupole deformations $\protect\beta_2$, and
the deformed single-particle orbitals occupied by the odd valence proton,
are compared with predictions of the macroscopic-microscopic mass model, and
with the experimental transition energies. All energies are in units of MeV;
the RHB spectroscopic factors are displayed in the sixth column.}
\label{TabA}
\begin{center}
\begin{tabular}{llllllllll}
& N & $S_p$ & $\beta_2$ & $p$-orbital & $u^2$ & $\Omega^{\pi}_p$ \cite
{MNK.97} & $S_p$ \cite{MNK.97} & $\beta_2$ \cite{MN.95} & $E_p$ exp. \\
\hline
$^{149}$Lu & 78 & -1.77 & -0.158 & $7/2^-[523]$ & 0.60 & $5/2^-$ & -1.51
& -0.175 & \\
$^{150}$Lu & 79 & -1.31 & -0.153 & $7/2^-[523]$ & 0.61 & $5/2^-$ & -1.00 &
-0.158 & 1.261(4) \cite{Sel.93} \\
$^{151}$Lu & 80 & -1.24 & -0.151 & $7/2^-[523]$ & 0.58 & $7/2^-$ & -0.99 &
-0.150 & 1.233(3) \cite{Sel.93}
\end{tabular}
\end{center}
\end{table}

\begin{table}
\caption{Spherical Ta ground-state proton emitters. RHB results for the 
proton separation energies, the single-particle orbitals occupied by the
odd proton, and the corresponding spectroscopic factors are compared 
with the predictions of the finite-range droplet (FRDM) mass model and 
with experimental data.}  
\label{TabB}
\begin{center}
\begin{tabular}{lllll}
&  & RHB/NL3 & FRDM \cite{MNK.97}  & EXP.  \\ 
\hline

$^{155}$Ta& S$_{p}$    & -1.677     &  -1.09     & -1.765(10)\cite{Uus.99} \\ 
            &J$^{\pi}$ & h 11/2$^{-}$ &  9/2$^{-}$ & 11/2$^{-}$     \\ 
            &spectr. factor      & 0.60       &            &   0.58(20)     \\
$^{156}$Ta &  S$_{p}$     & -1.129     &  -0.60  & -1.007(5)\cite{Page.92}   \\ 
 & J$^{\pi}$ &d 3/2$^{+}$ &  3/2$^{-}$   & 3/2$^{+}$\\ 
 &spectr. factor    &  0.51        &           &                  \\
$^{157}$Ta&        S$_{p}$      & -1.040   &  -0.48  & -0.927(7)\cite{Irv.97}\\ 
&  J$^{\pi}$  &h 11/2$^{-}$ &  9/2$^{-}$  & 1/2$^{+}$   \\ 
 &spectr. factor          & 0.42      &            &  0.56(24)           \\
\end{tabular}
\end{center}
\end{table}

\begin{references}
\bibitem{WD.97} P.J. Woods and C.N. Davids, Annu. Rev. Nucl. Part.
        Sci. {\bf 47}, 541 (1997).
\bibitem{ASN.97}  S. Aberg, P.B. Semmes, and W. Nazarewicz, Phys. Rev. C
                {\bf 56}, 1762 (1997).
\bibitem{Dav.98}  C.N. Davids {\it et al}, 
                Phys. Rev. Lett. {\bf 80}, 1849 (1998).
\bibitem{Ryk.99} K. Rykaczewski {\it et al},
                Phys. Rev. C {\bf 60}, 011301 (1999).
\bibitem{MFL.98}  E. Maglione, L.S. Ferreira, and R.J. Liotta, Phys. Rev.
Lett. {\bf 81}, 538 (1998);  Phys. Rev. C {\bf 59}, R589 (1999).
\bibitem{VLR.98} D. Vretenar,  G.A. Lalazissis, and P. Ring,
        Phys. Rev. C {\bf 57}, 3071 (1998).
\bibitem{LVR.99} G.A. Lalazissis, D. Vretenar,
        and P. Ring, Nucl. Phys. {\bf A650}, 133 (1999).
\bibitem{VLR.99} D. Vretenar,  G.A. Lalazissis, and P. Ring,
        Phys. Rev. Lett. {\bf 82}, 4595 (1999).
\bibitem{Uus.99} J. Uusitalo {\it et al},
        Phys. Rev. C {\bf 59}, R2975 (1999).
\bibitem{Bat.99} J. C. Batchelder {\it et al}, private communication.
\bibitem{LKR.97}  G. A. Lalazissis, J. K\"{o}nig and P. Ring; Phys. Rev.
        {\bf C55}, 540 (1997).
\bibitem{BGG.84}  J. F. Berger, M. Girod and D. Gogny; Nucl. Phys.
                 {\bf A428}, 32 (1984).
\bibitem{Sel.93} P.J. Sellin {\it et al},
        Phys. Rev. C {\bf 47}, 1933 (1993).
\bibitem{Page.92} R.D. Page {\it et al},
        Phys. Rev. Lett. {\bf 68}, 1287 (1992).
\bibitem{Irv.97} R.J. Irvine {\it et al},
        Phys. Rev. C {\bf 55}, R1621 (1997).
\bibitem{Hof.82} S. Hofman, W. Reisdorf, G. M\" unzenberg,
                F. P. Hessberger, J. R. H. Schneider, and 
                P. Armbruster, Z. Phys. A {\bf 305}, 125 (1982).
\bibitem{MNK.97}  P. M\"oller, J.R. Nix, and K.-L. Kratz, At. Data Nucl.
        Data Tables {\bf 66}, 131 (1997).
\bibitem{MN.95}  P. M\"oller, J.R. Nix, W.D. Myers, and W.J. Swiatecki, At.
        Data Nucl. Data Tables {\bf 59}, 185 (1995).
\bibitem{Liv.93} K. Livingston {\it et al},
                Phys. Rev. C {\bf 48}, R2151 (1993).


\end{references}
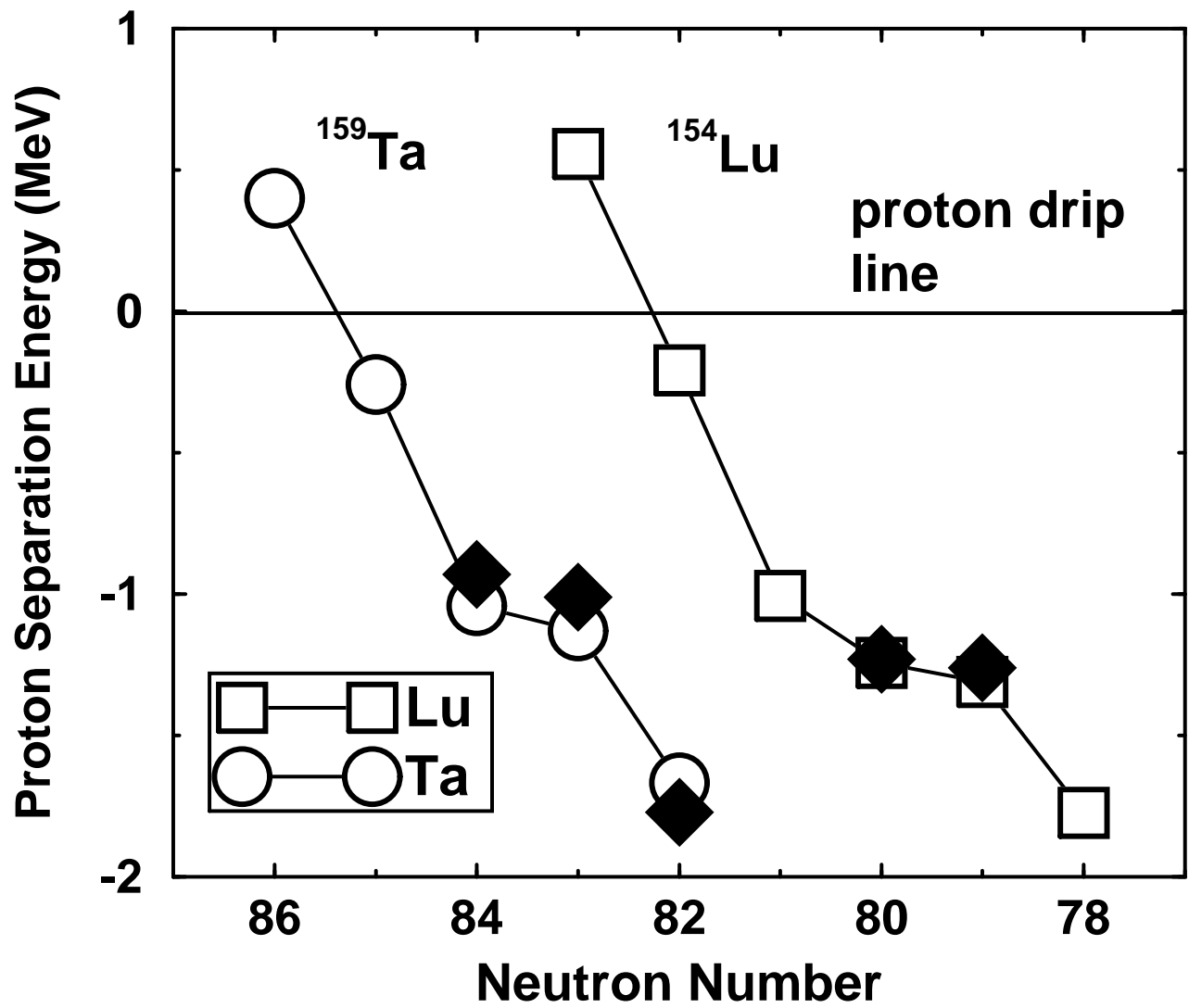
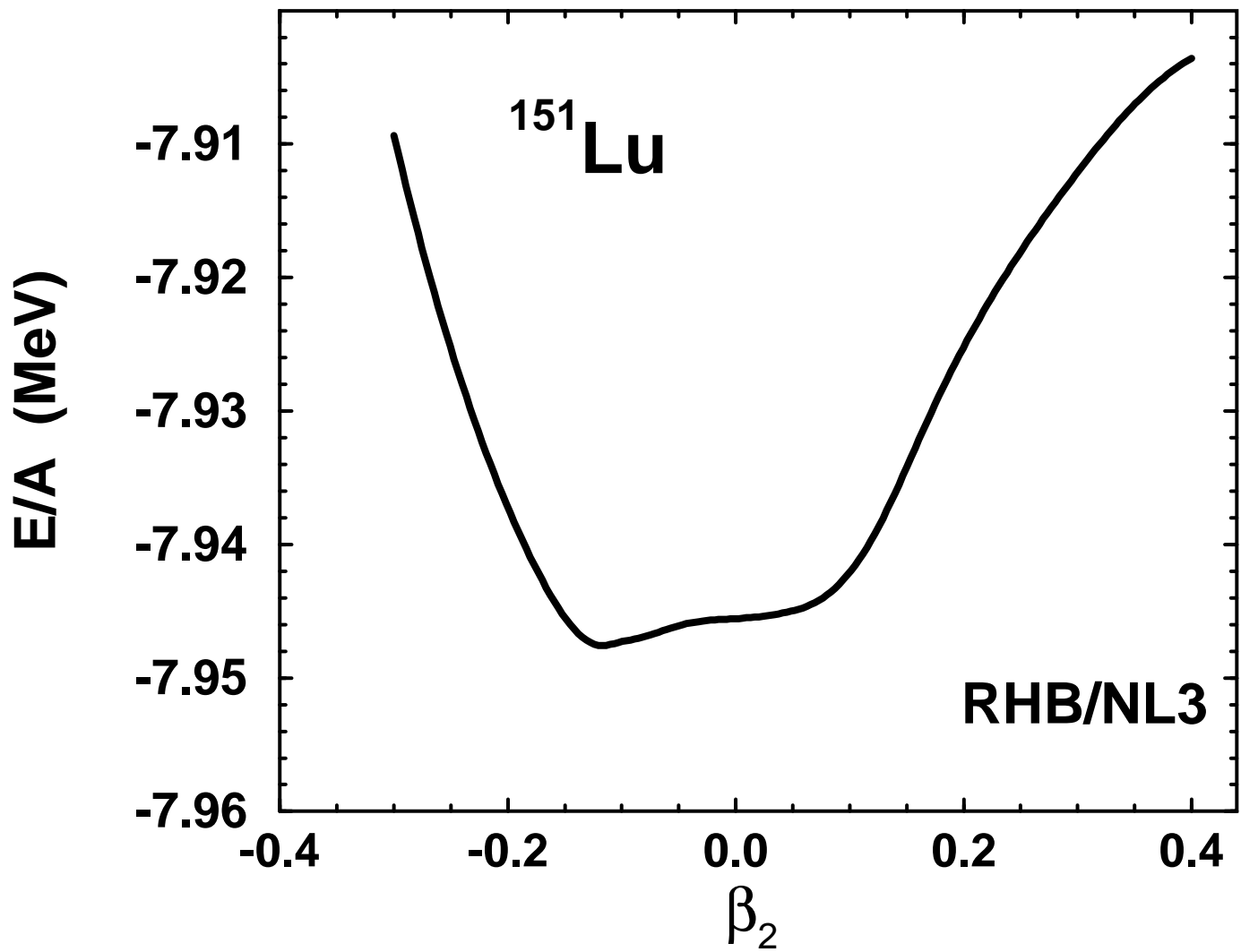
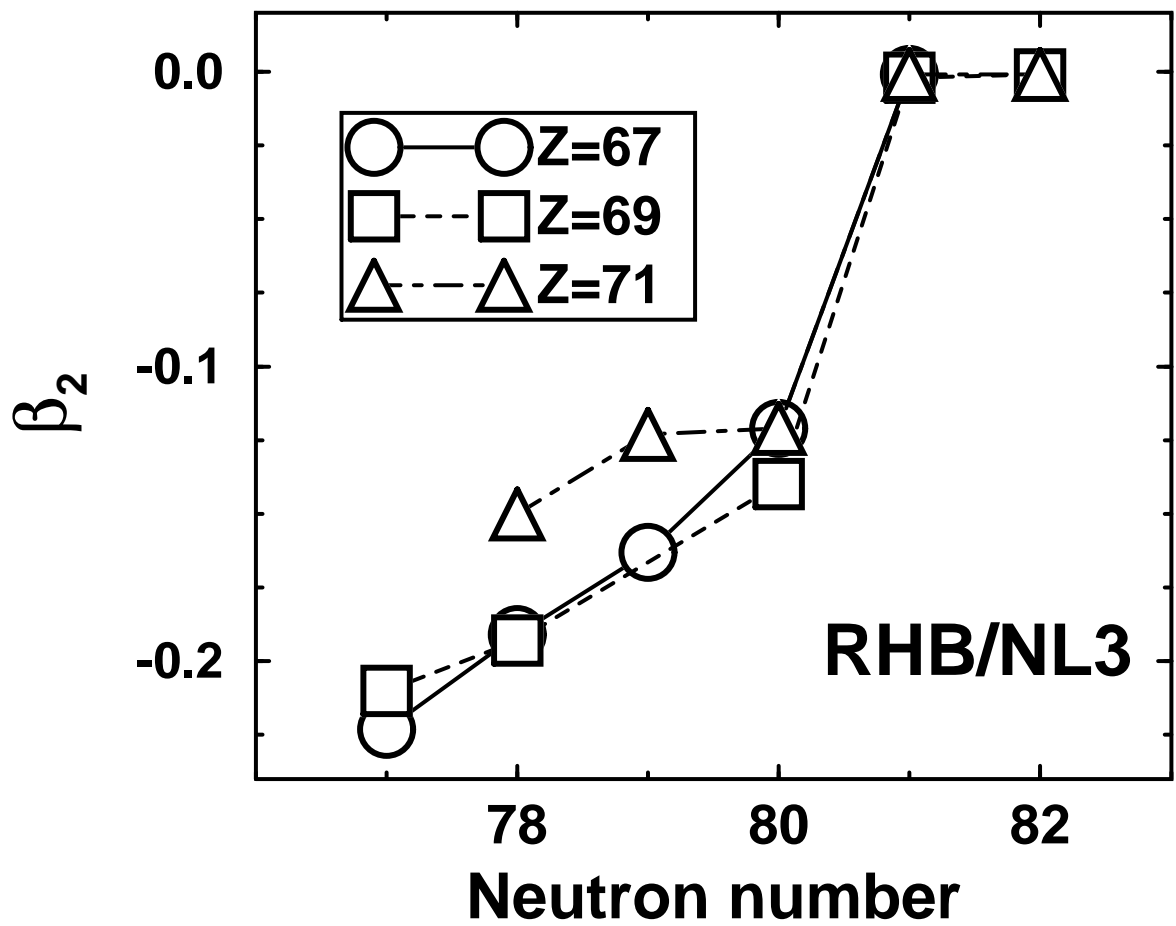
\end{document}